\documentclass[useAMS,usenatbib]{mn2e} 

\usepackage{graphicx} 
\usepackage{amsmath} 
\usepackage{epsfig}
\renewcommand{\ln}{\mathrm{ln}} 
\newcommand{\const}{\mathrm{const}} 
\renewcommand{\cos}{\mathrm{cos}}

\title[On the spherically-symmetric turbulent accretion] 
{On the spherically-symmetric turbulent accretion} 
\author[V. S. Beskin and K. V. Lezhnin]{V. S. Beskin$^{1, 2}$\thanks{E-mail: 
beskin@lpi.ru} and K. V. Lezhnin$^{2}$ 
\\ 
$^{1}$P.N.Lebedev Physical Institute, Leninsky prosp., 53, Moscow, 119991, Russia\\ 
$^{2}$Moscow Institute of Physics and Technology, Dolgoprudny, 
Moscow region, 141700, Russia} 

\begin{document} 

\date{Accepted, Received} 

\pagerange{\pageref{firstpage}--\pageref{lastpage}} \pubyear{2014} 

\maketitle 

\label{firstpage} 

\begin{abstract} 
We analize the quasi-spherical accretion in the presence of axisymmetric vortex 
turbulence. It is shown that in this case the turbulence changes mainly the effective 
gravity potential but not the effective pressure. 
\end{abstract} 

\section{Introduction}\label{aba:sec1} 

An activity of many astrophysical sources (Active Galactic Nuclei, Young Stellar Objects, 
Galactic X-ray sources, microquasars) is associated with the accretion. For this reason, 
the accretion onto compact objects (neutron stars or black holes) is the classical problem 
of modern astrophysics (see, e.g., \citet{ST, L} and references therein). At present the 
analytical approach, whose foundation was laid back in the mid-twentieth century~\citep{BH, B}, 
began to be supplanted by numerical simulations~\citep{Hunt, Petrich1, RA, T1, T2}. Analytical 
solutions were found only in exceptional cases~\citep{BK, Petrich2, And, BPi, BM, Pariev}. 

It should be emphasized that last time the focus of the research has been shifted to numerical
magnetohydrodynamic simulations, within which framework it has become possible to take into 
account the turbulent processes associated with magnetic reconnection, magnetorotational 
instability, etc.~\citep{BaHo, BS, KH}. However, in our opinion, some important features 
of the turbulent accretion can still be understood on the ground of simple analytical 
model. 

The problem of turbulent accretion and stellar wind has been discussed in various papers. 
\citet{Ax} showed that the inclusion of viscosity and heat conduction allows to remove 
singularity at the sonic surface and leads to the appearance of weak non-Rankine-Hugoniot 
shock waves in the solutions.~\citet{Koval} examined the spatial stability of spherical adiabatic 
Bondi accretion on to a point gravitating mass against external vortex perturbations.~\citet{BhRay} 
also discussed the role of turbulence in a spherically symmetric accreting system. In their paper 
it was shown that the sonic horizon of the transonic inflow solution is shifted inwards, in 
comparison with inviscid flow, as a consequence of dynamical scaling for sound propogation 
in accretion process.~\citet{Shch} formulated a model of spherically symmetric accretion flow 
in the presence of magnetohydrodynamic turbulence. 

In this paper we show on the ground of the simple model how the turbulence affects 
the structure of the spherically symmetric accretion. 
In the first part, we formulate the basic equations of ideal steady-state axisymmetric 
hydrodynamics, which are known to be reduced to one second-order equation for the stream 
function. Then, in the second part, the structure of the solitary curl is discussed in 
detail. Finally, in the third part we consider two toy models describing axisymmetric 
turbulence. It is shown that the turbulence changes mainly the effective gravity potential 
but not the effective pressure. 

\section{Basic equations} 

First of all, let us formulate basic hydrodynamical equations describing axisymmetric 
stationary flows. Then, as is well-known, it is convenient to introduce the potential 
$\Phi(r,\theta)$ connected with the poloidal velocity ${\bf v}_{\rm p}$ and the number 
density $n$ as~\citep{Heyvaerts-96, b3} 
\begin{equation} 
n{\bf v}_{\rm p} 
= \frac{\nabla \Phi \times {\bf e}_{\varphi}}{2\pi r\sin\theta}. 
\label{b1} 
\end{equation} 
This definition results in the following properties 
\begin{itemize} 
\item 
The continuity equation $\nabla \cdot (n {\bf v}) = 0$ is satisfied automatically. 
\item 
It is easy to verify that 
${\rm d} \Phi = n{\bf v} \cdot {\rm d} {\bf S}$, 
where ${\rm d}{\bf S}$ is an area element. As seen, the potential 
$\Phi(r,\theta)$ is a particle flux through the circle $r, \theta, 0<\varphi<2\pi$. In 
particular, the total flux through the surface of the sphere of radius $r$ is 
$\Phi_{\rm tot} = \Phi(r,\pi)$. 
\item 
As ${\bf v} \cdot \nabla \Phi = 0$, the velocity vectors ${\bf v}$ 
are located on the surfaces \mbox{$\Phi(r,\theta) = $ const.} 
\end{itemize} 
In this case, three conservation laws for energy $E_{\rm n}$, angular momentum 
$L_{\rm n}$, and the entropy $s$ can be formulated as 
\begin{eqnarray} 
E_{\rm n} & = & E_{\rm n}(\Phi) = \frac{{v}^2}{2} + {w} + \varphi_{\rm g}, 
\label{b2} \\ 
L_{\rm n} & = & L_{\rm n}(\Phi) = {v}_{\varphi}r\sin\theta, \\ 
s & = & s(\Phi). 
\label{b3} 
\end{eqnarray} 
Here $w$ is the specific enthalpy, and $\varphi_{\rm g}$ is the gravitational potential. 

In what follows we for simplicity consider the entropy $s(\Phi)$ to be constant. Then 
the equation for the stream function $\Phi(r,\theta)$ (which is no more than the 
projection of the Euler equation onto the axis perpendicular to the velocity vector 
${\bf v}$) looks like (cf.~\citet{Heyvaerts-96}) 
\begin{eqnarray} 
\varpi^2 \nabla_{k} 
\left(\frac{1}{\varpi^{2}n}\nabla^{k}\Phi\right) 
+4\pi^{2}nL_{\rm n}\frac{{\rm d}L_{\rm n}}{{\rm d}\Phi} 
-4\pi^{2}\varpi^{2}n\frac{{\rm d}E_{\rm n}}{{\rm d}\Phi}=0, 
\label{gscomp} 
\end{eqnarray} 
where $\varpi=r \sin \theta$. 
This equation represents the balance of forces in a normal direction to flow lines. In 
partucular, for spherically symmetric flow, i.e., for $E_{\rm n}(\Phi) =$ const, 
\mbox{$L_{\rm n}(\Phi) = 0$,} it has the solution 
\begin{equation} 
\Phi=\Phi_0(1-\cos~\theta). 
\label{gssol1} 
\end{equation} 

In the following, we deal with the linear angular operator 
\begin{equation} 
\hat{\cal L}_{\theta}=\sin\theta\frac{\partial}{\partial\theta}\left(\frac{1}{ 
\sin\theta} \frac{\partial}{\partial\theta}\right), 
\label{b22} 
\end{equation} 
originated from Eqn. (\ref{gscomp}). It has eigenfunctions 
\begin{eqnarray} 
& & Q_0=1-\cos~\theta, 
\label{b23} \\ 
& & Q_1=\sin^{2}\theta, 
\label{b24} \\ 
& & Q_2=\sin^{2}\theta~\cos~\theta, 
\label{b25} \\ 
& & \dots \nonumber \\ 
& & Q_m=\frac{2^{m}m!(m-1)!}{(2m)!} \sin^{2}\theta~ 
{\cal P}_m^{\prime}(\cos~\theta), 
\label{b26} 
\end{eqnarray} 
and the eigenvalues 
\begin{equation} 
q_m = -m(m+1). 
\label{b27} 
\end{equation} 
Here ${\cal P}_m(x)$ are the Legendre polynomials and the dash indicates their derivatives. 

Let us consider now the small disturbance of the spherically symmetric flow, so that 
one can write down the flux function as 
\begin{equation} 
\Phi=\Phi_0[1-\cos~\theta + \varepsilon^2 f(r, \theta)] 
\label{gssol1new} 
\end{equation} 
with the small parameter $\varepsilon \ll 1$. Then Eqn. (\ref{gscomp}) can be linearised, 
while the equation for the perturbation function $f(r,\theta)$ is written as~\citep{b3}: 
\begin{equation} 
\begin{aligned} 
-\varepsilon^2 D \frac {\partial^2 f} {\partial r^2} 
- \frac {\varepsilon ^2} {r^2} (D+1)\sin \theta \frac{\partial}{\partial \theta}(\frac{1}{\sin \theta} 
\frac{\partial f}{\partial \theta}) 
+\varepsilon^2 N_r \frac{\partial f}{\partial r}=\\ 
=-\frac{4\pi^2 n^2 r^2}{\Phi_0^2} \sin \theta (D+1) \frac{{\rm d} E_{\rm n}}{{\rm d} \theta} \\ 
+ \frac{4 \pi^2 n^2}{\Phi_0 ^ 2} (D+1) \frac{L_{\rm n}}{\sin \theta} \frac{d L_{\rm n}}{d \theta} 
-\frac{4\pi^2 n^2}{\Phi_0^2} \frac{\cos \theta}{\sin^2 \theta}L_{\rm n} ^2. 
\label{eqnlin} 
\end{aligned} 
\end{equation} 
Here $D=-1+c_{\rm s} ^2 / v^2$, and $N_r=2/r-4\pi^2 n^2 r^2 GM /\Phi_0^2$. 

This equation allows us to seek the solution in the form 
\begin{equation} 
f(r,\theta)=\sum_{m=0} ^{\infty} g_m (r) Q_m(\theta). 
\label {slnsum} 
\end{equation} 
Introducing now dimensionless variables 
\begin{equation} 
x=\frac{r}{r_*}, \, u=\frac{n}{n_*}, \, l=\frac{c_{\rm s} ^2}{c_* ^2}, 
\end{equation} 
where the $*$-values correspond to the sonic surface (which can be taken from the zero 
approximation), we can write the ordinary differential equations describing the radial 
functions $g_m (r)$: 
\begin{equation} 
\begin{aligned} 
(1-x^4 l u^2)g''_m 
+2\left(\frac{1}{x}-x^2 u^2\right)g'_m+m(m+1)x^2 l u^2 g_m=\\ 
=k_m \frac{R^2}{r_* ^2}x^4 l u^4 -\lambda _m \frac{R^2}{r_* ^2}u^2 - \sigma_m x^6 l u^4, 
\label{radialeq}. 
\end{aligned} 
\end{equation} 
Here $g'_m = {\rm d}g_m(x)/{\rm d}x$, $g''_m = {\rm d}^2g_m(x)/{\rm d}x^2$, 
and the expansion coefficients $\sigma_m$, $\lambda_m$ and $k_m$ depend on the disturbances 
as: 
\begin{equation} 
\sin \theta \frac{{\rm d}E_{\rm n}}{{\rm d}\theta} 
= \varepsilon^2 c_* ^2 \sum_{m=0} ^{\infty}\sigma_m Q_m(\theta), 
\label{denergy} 
\end{equation} 
\begin{equation} 
\frac{\cos~\theta}{\sin^2 \theta} L_{\rm n} ^2 
= \varepsilon^2 c_* ^2 r_* ^2 \sum_{m=0} ^{\infty}\lambda_m Q_m(\theta), 
\label{squaremomentum} 
\end{equation} 
\begin{equation} 
\frac{L_{\rm n}}{\sin \theta} \frac{{\rm d}L_{\rm n}}{{\rm d}\theta} 
= \varepsilon^2 c_* ^2 r_* ^2 \sum_{m=0} ^{\infty}k_m Q_m(\theta). 
\label{dmomentum} 
\end{equation} 

Finally, the functions $l(x)$ and $u(x)$ correspond to the spherically symmetric flow. 
For the polytropic equation of state $P(n,s)=A(s)n^{\Gamma-1}$ we use here they are connected 
by the relation $l=u^{\Gamma-1}$. As to the dimensionless number density $u(x)$, it can be 
found from ordinary differential equation 
\begin{equation} 
\frac{{\rm d} u}{{\rm d}x}=-2\frac{u}{x}\frac{(1-x^3 u^2)}{(1-x^4 l u^2)} 
\label{densityeq} 
\end{equation} 
with the boundary conditions 
\begin{equation} 
u(x)|_{x=1} = 1, 
\label{densityeqbound} 
\end{equation} 
\begin{equation} 
\left.\frac{{\rm d}u}{{\rm d}x}\right|_{x=1} = -\frac{4-\sqrt{10-6\Gamma}}{\Gamma+1}. 
\label{densityeqboundaries} 
\end{equation} 

\begin{figure} 
\includegraphics[scale=.25]{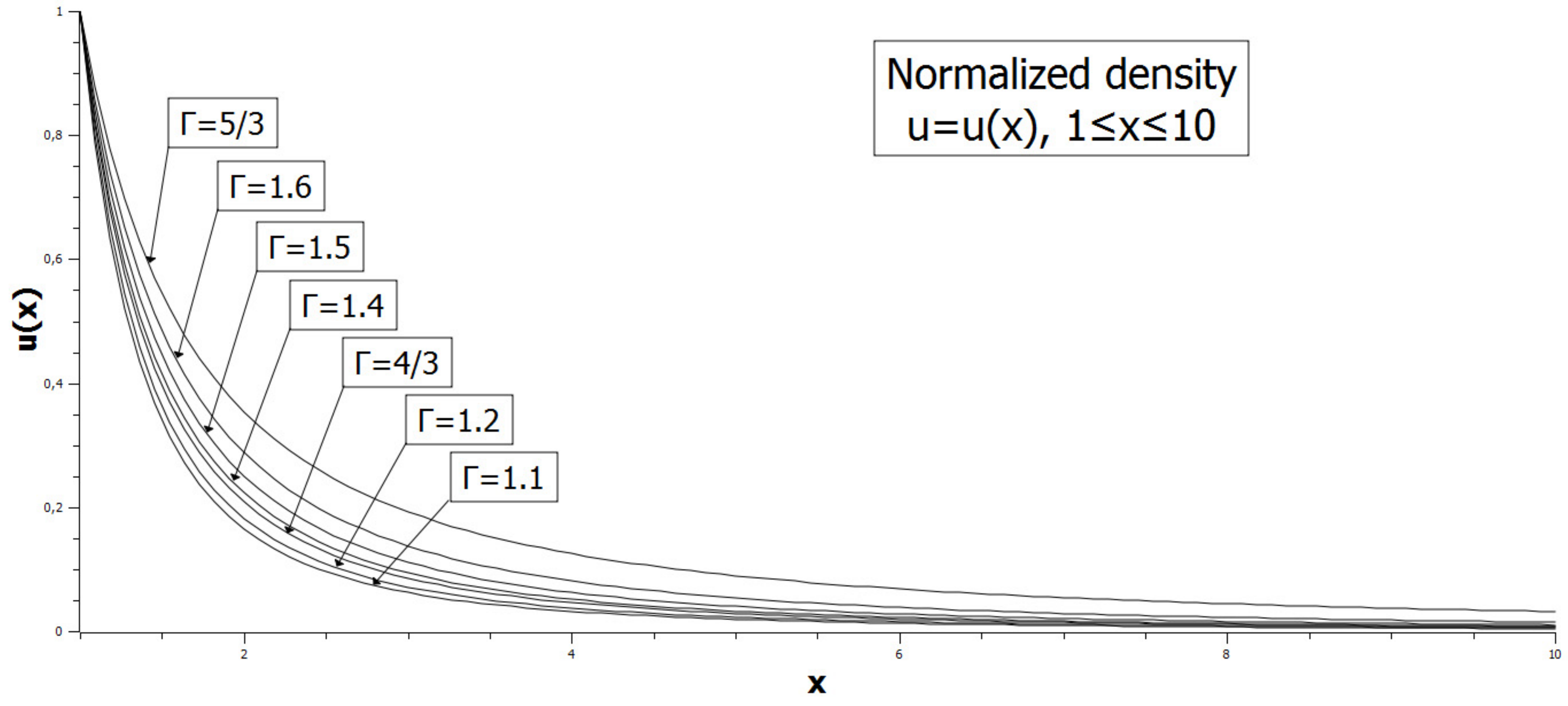} 
\caption{Dimensionless number density $u(x) = n/n_{*}$ in region $1\leq x \leq R/ r_*=10$, $\Gamma$ 
from 1.1 to 5/3} 
\end{figure} 

As for the boundary conditions for the system of Eqns. (\ref{radialeq}), they are taken 
analogous to the case of Bondi-Hoyle accretion~\citep{BPid}: 
\begin{equation} 
\varepsilon^2 g_m(1)=\frac{(2m)!}{2^m (m+1) m!} 
\left[ \frac{(\delta E_{\rm n})_m}{c_* ^ 2}-\frac{(L_ n ^2 /\sin^2 \theta)_m}{2c_* ^2 r_* ^2}\right], 
\label{boundg} 
\end{equation} 
\begin{equation} 
g' _m(R/r_*)=0, 
\label{boundderivg} 
\end{equation} 
where $(...)_m$ stands for the expansion in terms of the Legendre polynomials, 
which can be found from energy and angular momentum integral perturbations. 

\section{Solitary curl} 

Now let us consider in detail the internal structure of the quasi-spherical accretion with 
a small axisymetric perturbation localised near the axis $\theta = \pi$. In other words, we 
suppose that the angular size of the vortex is small enough ($\delta\theta_{\rm curl}\ll 1$). 
The main goal of this numerical calculation is to find the disturbance function $f(r, \theta)$ 
which gives us the possibility to determine velocity components $v_\theta$ and $v_{\varphi}$, 
i.e., the main characteristics of the perturbed flow. 

To model the internal structure of the vortex, we determine $\theta$-dependent angular velocity 
$\Omega(\theta)$ in the form 
\begin{equation} 
\Omega(\theta) = \Omega_0 \exp[-\alpha^2 (1+\cos~\theta)]. 
\label{momentum integral} 
\end{equation} 
Here $\Omega_0$ gives the amplitude of considered curl and the coefficient $\alpha \simeq 10$ 
(which are free parameters of our model) is an inversed curl width. Certainly, we assume that 
the perturbation is small in comparison with the main contribution of the radial accretion. 

Then the flow structure can be described by the system (\ref{radialeq}), 
(\ref{densityeq})--(\ref{boundderivg}) formulated in previous 
section. As to the expansion coefficients $k_m$, $\lambda_m$ and $\sigma_m$, they are to 
be determined from Eqns. (\ref{denergy})--(\ref{dmomentum}) and (\ref{boundg}) on the 
outer boundary of a flow $r = R$. For our choice (\ref{momentum integral}) the disturbances 
have the form 
\begin{equation} 
\delta E_{\rm n}(\theta) =\frac{{\Omega_0 ^2 \exp[-2\alpha^2 (1+\cos~\theta)] 
R^2 \sin^2\theta}}{2} + {o(\varepsilon^3)}, 
\label{energyperturb} 
\end{equation} 
\begin{equation} 
\delta L_{\rm n}(\theta) = {\Omega_0 \exp[-\alpha^2 (1+\cos~\theta)] R^2 \sin^2\theta}. 
\label{momentumperturb} 
\end{equation} 
As one can easily check, it gives $\varepsilon = \Omega_0 R/c_{\ast}$. 

Expansions (\ref{denergy}), (\ref{squaremomentum}), and (\ref{dmomentum}) in terms of 
$Q_m(\theta)$ contain some numerical difficulties because the set of this functions 
is not an orthogonal one, and, even though it converges, in our case of very small curl 
width we can neglect just summands with numbers larger than 50. Even in some trivial 
cases like $\Omega(\theta)\sim(1-\theta^2)$ these polynomials call a number of numerical 
obstacles (e.g., bad-conditioned matrix of linear equation for coefficients $k_m$, 
$\lambda_m$ and $\sigma_m$, etc). 

In order to expand functions of integrals, we used the auxiliary set of Chebyshev 
polynomials, which is orthogonal and posesses a feature of generally faster convergence. 
Using these polynomials, we could find all expansions with the accuracy no worse than 
$10^{-3}$. As was shown in Sect. 2, the normalized density function $u(x)$ can be derived 
from the equation (\ref{densityeq}) and boundary conditions (\ref{densityeqbound}) and 
(\ref{densityeqboundaries}). The results of numerical calculations for different polytropic 
indeces $\Gamma$ is shown on Figure 1. In particiular, as one can see, the density is nearly 
constant in subsonic regime ($r\gg r_\ast$). 

An example of the numerical calculation of perturbation function $f(r,\theta)$ is shown on 
Figure 2. We should stress that $f(r,\theta)$ turns actually zero outside the small region 
near the axial curl. This statement allows us to assume as a zero approximation that the 
turbulent accretion regime containing a number of curls can be considered as a set of 
noninteracting ones. Moving towards the accreting star, we can register that perturbation 
has a maximum on a radius $\sim 7 r_\ast$, then diminishes on the distance about $2 r_\ast$ 
and finally rapidly rises in the vicinity of the sonic surface $r_{\ast}$. 

Apart from numerical solution, we can also find the analytical asymptotic solutions in the 
supersonic region $x \ll 1$ where Eqn. (\ref{radialeq}) can be rewritten as 
\begin{equation} 
g''_m +\frac{3}{2x}g'_m+\frac{m(m+1)}{2^{\Gamma +1}}x^{-(3\Gamma+1)/2}g_m 
+\lambda_m \frac{R^2}{r_* ^2}\frac{1}{4x^3} = O(x^{-3}). 
\label{asympteq1} 
\end{equation} 
Here we take into account that $\Gamma < 5/3$. Getting rid of all parameters from the right part 
of this equation, one can introduce a new function $y_m=g_m/(\lambda_m R^2/r_*^2)$. Then 
Eqn. (\ref{asympteq1}) can be rewritten as 
\begin{equation} 
y''_m +\frac{3}{2x}y'_m+\frac{m(m+1)}{2^{\Gamma +1}}x^{-(3\Gamma+1)/2}y_m +\frac{1}{4x^3} 
= O(x^{-3}). 
\label{asympteq2} 
\end{equation} 
Neglecting now all terms which are proportional to $x^{-\nu}$, where $\nu > -3$, we obtain that 
this equation has an universal solution independent of the boundary conditions on the outer 
boundary $r = R$ 
\begin{equation} 
y(x)= -\frac{8}{x}.
\label{asymptotic} 
\end{equation} 
Remember that the same asymptotic behavior was obtained by~\cite{BM} for homogeneously rotating
flow.

\begin{figure} 
\includegraphics[scale=.30]{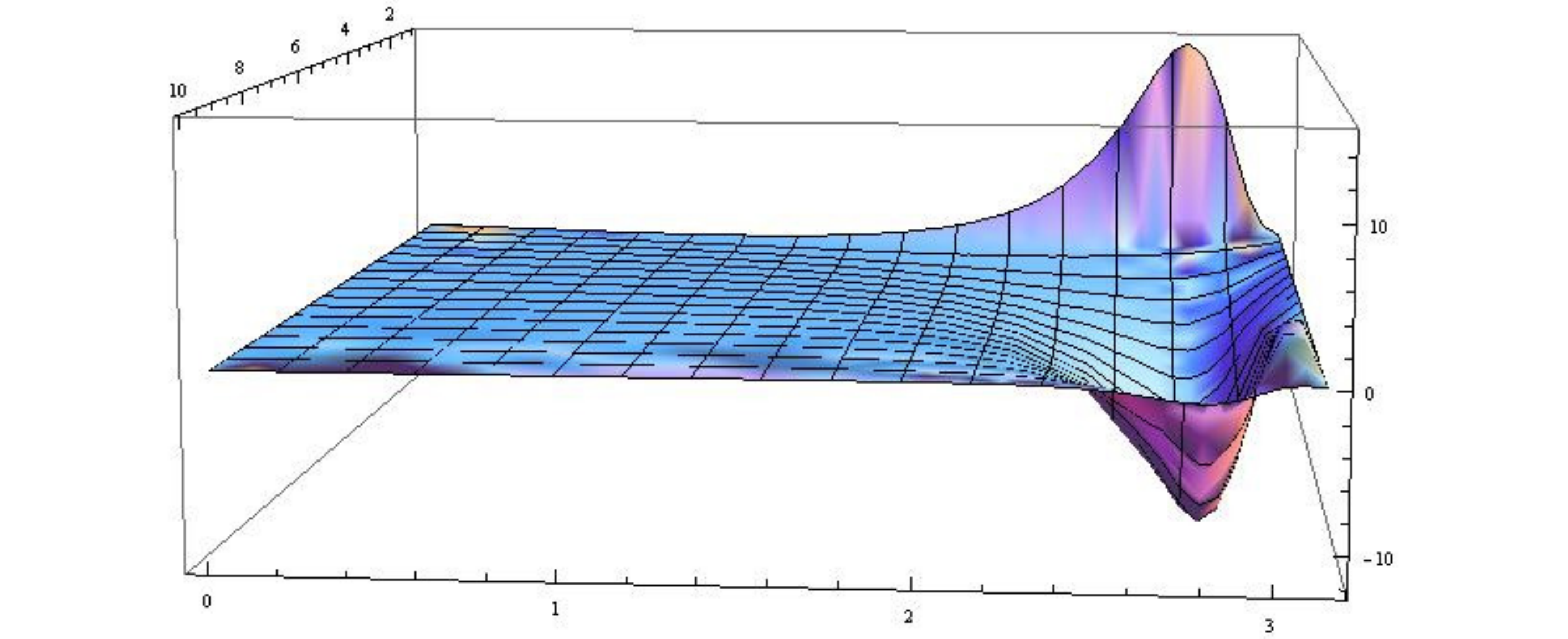} 
\caption{Disturbance function $f(r, \theta)$ in region $1\leq x \leq R/ r_*=10$, 
$0\leq \theta \leq \pi$, $\Gamma=4/3$} 
\end{figure} 

As was already stressed, numerical results allow us to determine $v_{\theta}/v_{\varphi}$ ratio 
around the curl. It is easy to show that 
\begin{equation} 
\frac{v_\theta}{v_\varphi}=2 \sqrt{2} \, \varepsilon \pi \left(\frac{r_*}{R}\right)^{7/2} p(r,\theta), 
\end{equation} 
where 
\begin{equation} 
p(r,\theta)=\frac{\displaystyle \sum_{m=0} ^{\infty} g'_{m} (r) Q_m(\theta)}{\displaystyle u(r) \sin^2\theta 
\exp[-\alpha^2(1+\cos~\theta)]}. 
\label{1} 
\end{equation} 
In our calculation we put $r_{*}/R = 0.1$, so that the function (\ref{1}) is limited in 
area near the curl ($|p(r,\theta)|<200$). Taking now into account that $\varepsilon$ is a small 
parameter of our expansion and $|f(r,\theta)|<20$, 
one can show that for reasonable parameter $\varepsilon$ the ratio $|v_{\theta}/v_{\varphi}|$ 
in the area of vortex has an order of $\sim 10 ^{-5}$ (see Figure 3). Thus, we could claim that 
$v_{\theta} \ll v_{\varphi}$, and then one can neglect the all terms in Navier-Stokes equations 
that consist $v_\theta$. 

\begin{figure} 
\includegraphics[scale=.70]{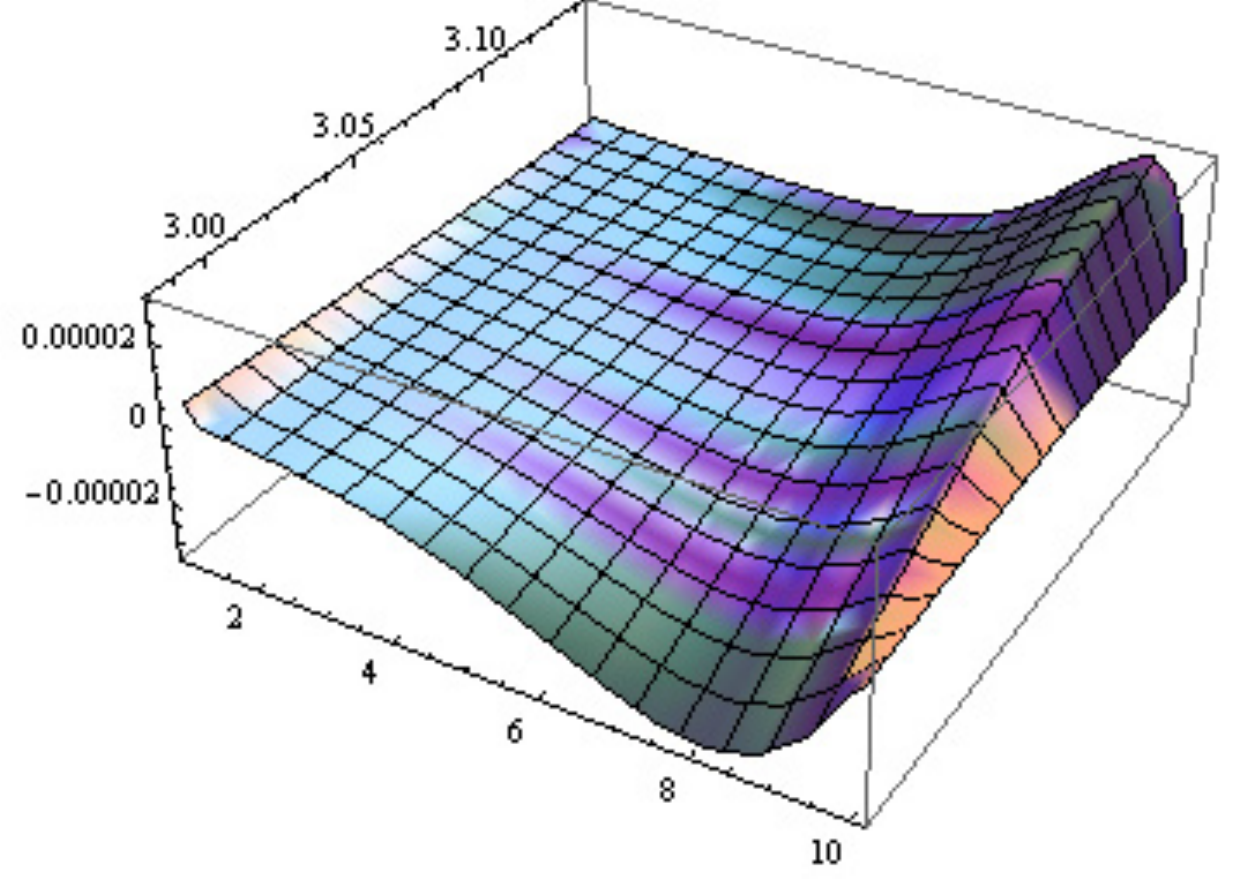} 
\caption{$v_\theta/v_{\varphi}$ ratio in region $1\leq x \leq10, 19\pi/20 \leq \theta \leq \pi$.} 
\end{figure} 

On the other hand, it is necessary to stress that, according to (\ref{asymptotic}), we cannot use 
our solution close enough to the origin ($r \rightarrow 0$). Indeed, analysing the field line 
equation $\displaystyle r \rm{d} \theta/\rm{d}r =\displaystyle v_\theta /v_r$, we obtain that the 
asymptotic solution survives until $\theta_0 \approx \delta \theta$, where $\theta_0 \ll 1$ is an 
initial angular size of a curl and $\delta \theta$ is a broadening parameter of the stream line. 
Deriving $\theta$-component of the velocity from (\ref{b1}) and taking $v_r$ of zero order, we get
\begin{equation}
\displaystyle \frac{\rm{d} \theta}{{\rm d}r} 
\sim \varepsilon ^2 \, \displaystyle \frac{\partial f / \partial r}{\sin \theta}.
\label{crit}
\end{equation}
Assuming now that $\theta \ll 1$, one can expand equation (\ref{crit}) in $\theta$ and neglecting 
all summands of $m$ with $m \geq 2$, we find
\begin{equation}
\displaystyle \frac{\rm{d} \theta}{{\rm d}r} 
\sim \displaystyle \varepsilon ^2 \theta \frac{\partial} {\partial r} \frac{R^2}{r r_{\ast}}.
\label{criteq}
\end{equation}
This equation can be simply integrated, and we obtain
\begin{equation}
\ln \frac{\theta_0+\delta \theta}{\theta_0}  \sim \varepsilon ^2 \frac{R^2}{r_\ast r}.
\end{equation}

Hence, under the radius $r \sim \varepsilon ^2 R^2 / r_\ast$  we cannot use the solution 
(\ref{asymptotic}) as the disturbance becomes larger than unity. In order to keep the solution 
up to star surface \mbox{$r = r_{\rm star}$,} we should demand 
\begin{equation}
\varepsilon ^2 R^2 / r_\ast < r_{\rm star},
\label{critansw}
\end{equation}
where $r_{\rm star}$ corresponds to the star radius. It gives us the general condition of 
appliability of the approach described above. Unless we cannot use the method of linear expansion 
of Grad-Shafranov equation, and the turbulent flow is to be described in another way which lies 
outside the consideration of current paper. 

Let us try now, as have been done in some papers (see, e.g.,~\cite{Shaq}, \cite{Sharura}), to describe 
the vortex disturbance through some effective pressure, i.e., to calculate a small correction for the 
pressure function $P$ caused by the presence of turbulent curl. Starting with $\theta$-component 
of Euler equation 
\begin{equation} 
v_r \frac{\partial v_\theta}{\partial r}+v_\theta\frac{1}{r} 
\frac{\partial v_\theta}{\partial \theta}+\frac{v_r v_\theta}{r}-\frac{v_{\varphi} ^2}{r} 
\cot\theta = -\frac{1}{r}\frac{\partial P / \partial \theta}{\rho}
\label{euler_theta} 
\end{equation} 
and neglecting all the terms containing $v_\theta$, we obtain 
\begin{equation} 
-\frac{v_{\varphi} ^2}{r} \cot \theta=-\frac{1}{r}\frac{\partial P / \partial \theta}{\rho}. 
\end{equation} 
Expanding this equation near the axis and assuming 
\begin{equation} 
v_{\varphi}=\frac{L}{r \sin \theta} = \frac{K \theta}{r}, 
\end{equation} 
where $K =$ const, we obtain 
\begin{equation} 
\rho \frac{K^2 \theta}{r^2} = \frac{\partial P }{ \partial \theta}.
\end{equation} 
It gives for the additional pressure $P_{\rm add}$ 
\begin{equation} 
P_{\rm add} \sim \frac{\rho K^2}{\alpha^2 r^2}, 
\end{equation} 
where $\alpha^{-1}$ is an approximate size of a curl. 

Thus, one can conclude that the pressure $P_{\rm add}$ has explicit dependence on $r$, 
which cannot be taken into consideration using any equation of state. Hence, we should 
include the turbulence into the consideration in another way than introducing effective 
turbulent pressure as often suggested by~\citet{Shaq, Sharura}. 

\section{Two toy models} 

Let us suppose now that the turbulence in the accreting matter can be described by the 
large number of axisymmetric vortexes with different parameters $\Omega_{0}$ and $\alpha$ 
filling all the accreting volume. Within this approach one can construct two toy analytical 
models demonstrating how the turbulence can affect the structure of the spherically symmetric 
accretion. 

\subsection{Inviscous flow} 

The first model in which we neglect viscosity corresponds to classical ideal spherically 
symmetric Bondi accretion~\citep{B}. 
In this case one can consider the following system of equations: 
\begin{eqnarray} 
\nabla (n {\bf v}) & = & 0, \\ 
({\bf v} \cdot \nabla){\bf v} & = & -\frac{\nabla P}{\rho}-\nabla \varphi_{\rm g}, \\ 
({\bf v} \cdot \nabla) s & = & 0. 
\end{eqnarray} 
Following~\citet{B} we consider polytropic equation of state $P = P(n,s) = k(s)n^\Gamma$ 
resulting in for polytropic index $\Gamma \neq 1$ 
\begin{eqnarray} 
c_{\rm s} ^2 & = & \frac{\Gamma k}{m_p}n^{\Gamma-1}, \\ 
w & = & \frac{c_{\rm s} ^2}{\Gamma-1}, \\ 
T & = & \frac{m_p}{\Gamma}c_{\rm s} ^2 . 
\end{eqnarray} 
Here again $n$ ($1/{\rm cm}^{3}$) is the number density, $m_{\rm p}$ (in ${\rm g}$) is 
the mass of particles ($\varrho = m_{\rm p}n$ is the mass density), 
$s$ is the entropy per one particle (dimensionless), 
\hbox{${w}$ (in ${\rm cm}^2/{\rm s}^2$)} is the specific enthalpy, 
$T $ (in ${\rm erg}$) is the temperature in energy units, 
and, finally, $c_{\rm s}$ (${\rm cm/s}$) is the sound velocity. 

As was demonstrated above, for a weak enough turbulence level (\ref{critansw}) for any 
inividual curl one can neglect $\theta$-component of the velocity perturbation in 
comparison with the toroidal one $v_{\varphi}$ up to the central body $r = r_{\rm star}$. 
Thus, in zero approximation one can put $v_{\theta} = 0$, i.e., $\theta =$ const. This 
implies that, according to the angular momentum concervation law $r \sin\theta~v_{\varphi} =$ 
const, we can write down 
\begin{equation} 
v_{\varphi}(r, \theta) = \Omega(\theta)\frac{R^2}{r}\sin\theta. 
\label{ansatz} 
\end{equation} 
Here $\Omega(\theta)$ is a smooth function of $\theta$ that can be approximately 
described as: 
\begin{eqnarray} 
\Omega(\theta) \approx \begin{cases} \Omega_{0}, & \pi - \alpha^{-1} < \theta < \pi, \\ 
0 & 0 < \theta < \pi - \alpha^{-1}. \end{cases} 
\end{eqnarray} 

In order to find the characteristic values of the accretion flow we have to use energy 
and momentum intergals conserving on stream lines. Taking into account an assertion  
$v_r \gg v_\varphi \gg v_\theta$, we can neglect $\theta$-component of velocity which 
gives 
\begin{eqnarray} 
E_{\rm n}(\theta) & = & \frac{v_r ^2 (r)}{2}+\omega(r) 
+\varphi_{\rm g} (r)+\frac{L^2 (\theta)}{2 r^2 \sin ^2 \theta},\\ 
L_{\rm n}(\theta) & = & v_{\varphi} r \sin \theta=\Omega (\theta) r^2 \sin ^2 \theta . 
\end{eqnarray} 
Averaging now these integrals in $\theta$ and introducing a new value 
\begin{equation} 
L_{\rm av}^2 \equiv \langle \frac{L^2 (\theta)}{\sin ^2 \theta} \rangle 
\label{AveMomentum} 
\end{equation} 
we obtain for the averaged energy integral 
\begin{equation} 
E_{\rm av}\equiv \langle E_{\rm n}(\theta) \rangle 
=\frac{v_r ^2 (r)}{2} +\omega(r) 
+\underbrace{\varphi_{\rm g} (r) +\frac{L_{\rm av}^2}{2 r^2}}_{\varphi_{\rm eff}(r)}. 
\label{E_av} 
\end{equation} 
As we see, two last terms can be considered as effective gravitational potential. 

Futher calculations are quite similar to the classical Bondi problem for the spherical 
flow. In other words, using another integrals of motion, i.e., the total particle flux 
\begin{equation} 
\Phi= 4\pi r^2 n(r) v_r (r)= \const, 
\end{equation} 
and the entropy $s$, one can rewrite the energy integral (\ref{E_av}) as 
\begin{equation} 
E_{\rm av}=\frac{\Phi ^2}{32 \pi ^2 n^2 r^4}+\frac{\Gamma k(s)}{\Gamma-1} 
\frac{n^{\Gamma -1}}{m_p}-\frac{GM}{r}+\frac{L_{\rm av} ^2}{2 r^2}. 
\end{equation} 
It gives the following expression for the logarithmic $r$-derivative of 
the number density 
\begin{equation} 
\eta_1 = \frac{r}{n} \frac{{\rm d} n}{{\rm d} r} 
= \frac{\displaystyle 2-\frac{GM}{v_r ^2 r} 
+\frac{L_{\rm av} ^2}{v_r ^2 r^2}}{\displaystyle -1+\frac{c_{\rm s} ^2}{v_r ^2}}. 
\end{equation} 
As for Bondi accretion, this derivative has a singularity on the sonic surface 
${v_r} = c_{\rm s} = c_{\ast}$. This implies that for smooth transition through 
the sonic surface $r=r_{\ast}$, the additional condition is to be satisfied: 
\begin{equation} 
2-\frac{GM}{c_\ast ^2 r_\ast}+\frac{L_{\rm av} ^2}{c_\ast ^2 r_\ast^2}=0. 
\label{bondieq}
\end{equation} 

Solving now (\ref{bondieq}) in terms of $r_\ast$ in this approximation, we find 
\begin{eqnarray} 
r_* & = & \frac{GM}{2c_* ^2}\left(1-\frac{4 L_{\rm av}^2 c_* ^2}{G^2M^2}\right),\\ 
c_* & = & \sqrt{\frac{2}{5-3\Gamma}}c_{\infty}\left(1+\frac{12 (\Gamma-1)}{(5-3\Gamma)^2} 
\, \frac{L_{\rm av}^2 c_{\infty}^2}{G^2M^2}\right), 
\label{sound1} 
\end{eqnarray} 
where $c_\infty$ is evaluated from 
\begin{equation} 
E_{\rm av}=\frac{c_\infty ^2}{\Gamma -1}. 
\end{equation} 
Accordingly, we obtain for
$r_\ast / r_{\ast \rm{B}}$ and $c_\ast / c_{\ast \rm{B}}$ ratios: 
\begin{eqnarray} 
\frac{r_\ast }{r_{\ast \rm{B}}} & = & 1 - \frac{16}{(5-3\Gamma)^2} 
\, \frac{L_{\rm av}^2 c_{\infty}^2}{G^2M^2} , \\ 
\frac{c_{\ast} }{ c_{\ast \rm{B}}} & = & 1 + \frac{12 (\Gamma-1)}{(5-3\Gamma)^2} 
\, \frac{L_{\rm av}^2 c_{\infty}^2}{G^2M^2}, 
\end{eqnarray} 
where $c_{\ast \rm{B}}$ and $r_{\ast \rm{B}}$ correspond to the classical Bondi accretion. 
Finally, using the definition (\ref{AveMomentum}) for $L^2 _{\rm av}$, we can rewrite 
our criteria of the appliability (\ref{critansw}) as $L^2_{\rm av}  \ll GM r_{\rm star}$. 
As $r_{\rm star}<r_\ast$, it can be finally rewritten as  
\begin{equation}
L_{\rm av} \ll \frac{GM}{c_\ast}.
\end{equation}
 
To sum up, one can conclude that the nonzero angular momentum effectively decreases the 
gravitational force. In other words, the presence of the angular momentum do not allow matter 
to fall down as easy as in its absence. Roughly speaking, we substitute our gravitating 
center with one that posesses less mass. So, in the case of Bondi accretion with a small 
angular momentum perturbation we should modify the relations for sonic surface radius and 
velocity --- they decrease and rise respectively. It is important to note that we can 
consider turbulent accretion regime as one with a modified gravity potential. 

\subsection{Viscous flow} 

\begin{figure} 
\includegraphics[scale=.24]{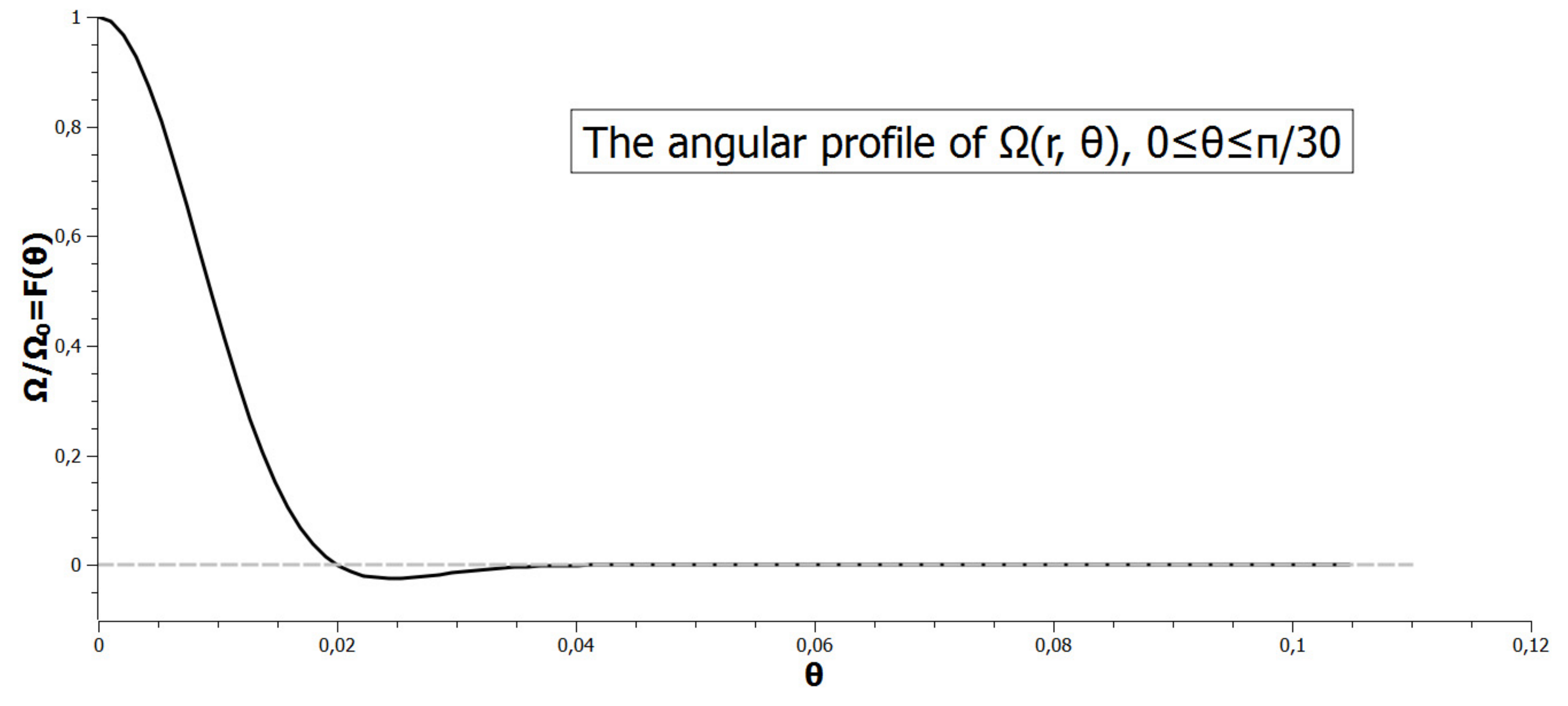} 
\caption{Ratio $\Omega / \Omega_0=F(\theta)$ for
$0\leq \theta \leq \pi/30$} 
\end{figure} 

It this subsection we consider stationary axisymmetric quasi-spherical flow of viscous 
fluid. Using the condition $v_r \gg v_\varphi \gg v_\theta$, and neglecting all the
terms containing $v_\theta$, we obtain for $\varphi$-component of Euler equation~\citep{LL}
\begin{equation} 
v_r \frac{\partial v_{\varphi}}{\partial r} + \frac{v_r v_\varphi}{r} 
=\nu \left(\nabla^2 v_\varphi -\frac{v_\varphi}{r^2 \sin ^2 \theta}\right). 
\label{LL} 
\end{equation} 
For viscous flow it is convenient to determine the toroidal component of the velocity 
$v_\varphi$ as 
\begin{equation} 
v_\varphi=\Omega(r,\theta) r \sin \theta, 
\end{equation} 
where we will use the following form for the angular velocity $\Omega (r, \theta)$: 
\begin{equation} 
\Omega(r,\theta)=\Omega_0 (r) \exp \left[-\frac{\theta^2}{2 \delta(r)}\right]. 
\label{Om} 
\end{equation} 
Here $\Omega_0=\Omega_0(r)$ is an amplitude, and $\delta=\delta(r)$ is a square of effective angular 
width of an individual curl. Substituting now $v_\varphi$ into Eqn. (\ref{LL}), we obtain 
\begin{equation} 
\dot{M} \frac{\rm d}{{\rm d}r}(\Omega r^2 \sin \theta) = \frac{4 \pi r^2 \eta}{\sin ^2 \theta} 
\frac{\rm d}{{\rm d} \theta}\left[\sin ^3 \theta \frac{{\rm d} \Omega}{{\rm d} \theta}\right], 
\label{54} 
\end{equation} 
where 
\begin{equation} 
\dot{M}=4 \pi r^2 \rho v_r 
\label{accretionrate} 
\end{equation} 
is the accretion rate remaining constant in stationary flow, and $\eta = \rho \nu$ is a dynamic 
viscous coefficient which can be considered as a constant as well~\citep{LL}. 

Using now Eqn. (\ref{54}), one can easily show that the total angular momentum of an individual 
vortex conserves (${\rm d } L/ {\rm d}r=0$). Indeed, internal friction connecting with viscosity
cannot change the total angular momentum of the accreting matter. For this reason, together with
(\ref{accretionrate}), the angular momentum
\begin{equation} 
{\rm d}L = \rho \Omega r^2 \sin ^2 \theta {\rm d} \phi \sin \theta {\rm d} \theta r^2 {\rm d}r 
\label{56} 
\end{equation} 
can be rewritten as a full $\theta$-derivative. This implies that the r.h.s. of Eqn. (\ref{56}) 
intergated over $\theta$ becomes zero. 

Further, to determine the radial dependence of the curl amplitude $\Omega_{0}(r)$ and the 
squared width $\delta(r)$, we substitute the angular velocity $\Omega_{0}(r,\theta)$ (\ref{Om}) 
into (\ref{LL}) and expand it in terms of $\theta$ near the axis, neglecting all the terms with 
the power more than 3. As a result, we obtain two equations for $\Omega_{0}(r)$ and $\delta(r)$ 
\begin{equation} 
\frac{\dot{M}}{2 \pi \eta} r \Omega_0 (r) 
+\frac{4r^2 \Omega_0 (r)}{\delta(r)}+\frac{\dot{M}}{4 \pi \eta} r^2 \Omega_0 ' (r)=0, 
\end{equation} 
\begin{equation} 
\begin{split} 
-18r^2 \Omega_0 (r) -\frac{3\dot{M}}{2 \pi \eta }r \delta (r) \Omega_0 (r)-10r^2 \delta (r) 
\Omega_0 (r)- \\ 
\frac{\dot{M}}{2 \pi \eta}r \delta ^2 (r) \Omega_0 (r) 
+\frac{3 \dot{M}}{4 \pi \eta}r^2 \delta'(r) \Omega_0 (r) -\frac{\dot{M}}{4 \pi \eta}r^2 
\delta ^2 (r) \Omega_0 ' (r)=0, 
\end{split} 
\end{equation} 
which have simple solutions 
\begin{eqnarray} 
\delta (r) & = & \delta_0+\frac{8 \pi \eta(r-r_{0})}{\dot{M}},\\ 
\Omega_0 (r) & = & \Omega_0 \frac{r_0^2}{r^2} \left[\frac{8 \pi \eta (r-r_{0})}{\dot{M} \delta_0} 
+1\right]^{-2}. 
\end{eqnarray} 
Introduction of a small vortex tubulence can be again treated by modifying a 
gravitational potential as
\begin{equation}
\varphi_{\rm{eff}} = \displaystyle - \frac{GM}{r}+\frac{\Omega_0 ^2 r_0 ^4 \delta_0}{4 \pi r^2} 
\left[\displaystyle 1+\frac{8 \pi \eta}{\dot{M}\delta_0}(r-r_0) \right]^{-3}.
\end{equation}

Thus, viscosity results in increasing of the vortex width ($\delta^{\prime} < 0$ for
${\dot M < 0}$ corresponding to accretion) and diminishing of the angular rotation. 
On the other hand, the role of viscosity will be small if
\begin{equation} 
\frac{8 \pi \eta r_0}{|\dot{M}|\delta_{0}}\ll 1. 
\label{Remm}
\end{equation} 
For $\eta \rightarrow 0$ we return to the previous result $\delta(r) =$ const, 
$\Omega_0(r) \propto r^{-2}$. Introducing now Reynolds number as 
\begin{equation} 
{\rm Re }= \frac{\rho v l}{\eta}, 
\end{equation} 
where $\rho, v$ and $l = r_0 \delta^{1/2}$ are characteristic values of a flow and 
using expression (\ref{accretionrate}), we can rewrite (\ref{Remm}) as 
\begin{equation} 
{\rm Re }= \frac{|\dot{M}|}{4 \pi r_0 \eta} \gg \delta_0 ^{-1/2}. 
\end{equation} 
This implies that the role of viscosity is small for turbulent flow. 


Certainly, the analysis presented above allows us to take into consideration only isolated 
set of curls. In reality, dense celluar turbulent structure posesses a number of collective 
effects~\citep{ProkhPopov}, that is to be described in another way. The easiest method to 
proceed with the minimal number of additional assumptions is to choose another angular 
velocity profile.

As the total angular momentum of the accreting matter is suppose to be zero, we will use the 
following expression for the angular velocity 
\begin{equation} 
\Omega(r, \theta)=\Omega_0 (r) \exp \left(-\frac{\theta^2}{2 \delta}\right)
\left(1-\frac{\gamma^2}{2 \delta}\theta ^2\right).
\end{equation} 
Here the parameter $\gamma$ is to be chosen from the condition of the zero total angular momentum 
\begin{equation} 
\int_{|\vec{r}| \leq R} {\rm d} L=0, 
\end{equation} 
which is equivalent to 
\begin{equation} 
\int_{0} ^ {\pi} d\theta \sin^3 \theta \, \Omega(r,\theta)=0. 
\end{equation} 
One of its realisations can be seen on Figure 4 where the dashed line shows zero 
angular velocity level. 
Configuration like this one represents the unit of celluar turbulent structure, fulfilling main 
requirements of its nature. To simplify our calculations, we hold $\gamma$ and $\delta$ on constant 
values in order to get simple equation on $\Omega_0 (r)$. Again, we expand equation (\ref{54}) in 
terms of $\theta$ to the first order. As a result, we obtain for $\Omega_0(r)$: 
\begin{equation} 
\Omega_0 (r)=\Omega_0 \left(\frac{r_0}{r} \right)^2 
\exp \left[\frac{16 \pi \eta}{\dot{M}}\frac{(1+\gamma^2)}{\delta}(r_0 -r)\right]. 
\label{eff3}
\end{equation} 
In this case, the effective gravitational potential cannot be derived for arbitary parameters 
without special functions. It can be written as
\begin{equation}
\varphi_{\rm{eff}} = \displaystyle 
- \frac{GM}{r}+C \cdot \frac{\displaystyle \Omega_0 ^2 r_0 ^4}{\displaystyle r^2} 
\exp \left[ - \frac{16 \pi \eta}{|\dot{M}|}\frac{(1+\gamma^2)}{\delta}(r_0 - r) \right],
\end{equation} 
where 
\begin{equation}
\displaystyle C=\frac{1}{2 \pi}\int_0 ^\pi\rm{ d}\theta 
\exp(-\theta^2/\delta)(1-\frac{\gamma ^2 \theta^2}{ 2  \delta})^2 \sin^2 \theta. 
\end{equation}
Choosing, for instanse, $\gamma=1/\sqrt{2}$ and $\delta = 10^{-4}$, it gives 
$C\approx3.4 \cdot 10^{-8}$.

The expression under the exponential function in (\ref{eff3}) is always lower than zero, so the 
criterion of the importance of viscosity effects can be formulated as 
\begin{equation} 
\frac{16 \pi \eta}{|\dot{M}|}\frac{(1+\gamma^2)}{\delta}r_0 \gg 1, 
\end{equation} 
which is more convinient to discuss in terms of Reynolds number 
\begin{equation} 
\delta^{-1/2}\gg {\rm Re}. 
\end{equation} 
Thus, for narrow vortex (i.e., for $\delta<10^{-2}$) the viscosity effects must be taken into 
account. 

\section{Conclusion} 

We have shown that in the case of the vortex turbulence represented by a solitary axisymmetric 
curl, one cannot simulate effects of the turbulence by introduction of the effective pressure 
which is common in recent papers (see, e.g.,~\citet{Shaq}, \citet{Sharura}). On the other hand, 
it is possible to take into account the vortex structure of the accreting flow by introducing an 
effective gravitational potential. 


Further, we described two analytical toy models that show how the turbulence affects the 
structure of the spherically symmetric flow. In particular, it was shown that the sonic 
surface moves inwards because of effective diminishing of gravitational force. Finally, 
a criterion to analyze the importance of viscosity effects in the adiabatic flow filled 
either by isolated or dense set of curls was formulated. 

We thank K.P.~Zybin for useful discussions. This work was partially 
supported by Russian Foundation for Basic Research (Grant no.~14-02-00831) 

{} 

\end{document}